# Writing on Nanocrystals: Patterning Colloidal Inorganic Nanocrystal Films through Irradiation-Induced Chemical Transformations of Surface Ligands

Francisco Palazon† , Mirko Prato‡ , and Liberato Manna*†

†Nanochemistry Department and ‡Materials Characterization Facility, Istituto Italiano di Tecnologia, Via Morego 30, 16163 Genova, Italy

## Abstract

In the past couple of decades, colloidal inorganic nanocrystals (NCs) and, more specifically, semiconductor quantum dots (QDs) have emerged as crucial materials for the development of nanoscience and nanotechnology, with applications in very diverse areas such as optoelectronics and biotechnology. Films made of inorganic NCs deposited on a substrate can be patterned by e-beam lithography, altering the structure of their capping ligands and thus allowing exposed areas to remain on the substrate while non-exposed areas are redispersed in a solvent, as in a standard lift-off process. This methodology can be described as a “direct” lithography process, since the exposure is performed directly on the material of interest, in contrast with conventional lithography which uses a polymeric resist as a mask for subsequent material deposition (or etching). A few reports from the late 1990s and early 2000s used such direct lithography to fabricate electrical wires from metallic NCs. However, the poor conductivity obtained through this process hindered the widespread use of the technique. In the early 2010s, the same method was used to define fluorescent patterns on QD films, allowing for further applications in biosensing. For the past 2–3 years, direct lithography on NC films with e-beams and X-rays has gone through an important development as it has been demonstrated that it can tune further transformations on the NCs, leading to more complex patternings and opening a whole new set of possible applications. This Perspective summarizes the findings of the past 20 years on direct lithography on NC films with a focus on the latest developments on QDs from 2014 and provides different potential future outcomes of this promising technique.

## Introduction

The ability to fabricate monodisperse inorganic nanocrystals (NCs) on large scales and under relatively mild conditions by colloidal syntheses has been one of the main driving forces in the development of nanoscience and nanotechnology in the past couple of decades.(1, 2) Numerous reports on these syntheses have been published from the seminal works of the early 1980s(3-12)to the present time. Beyond the simple “dot” shape, anisotropic NCs are commonly synthesized, including 2D structures (nanoplatelets and nanosheets)(13-19) and 1D structures (nanorods and nanowires).(20-24) Furthermore, complex geometries including branched and core–shell NCs have also been demonstrated.(25-29) Other than the shape diversity, colloidal NCs may be formed from different chemical compositions including metals(30-32) and oxides(33-36) as well as binary, ternary, and quaternary semiconductors,(37-47) more commonly referred to as quantum dots (QDs). This diversity directly translates into a multitude of physical properties that are unique to inorganic NCs, among which we can cite superparamagnetism,(48, 49) high catalytic activity,(50-

54) support of localized surface plasmons,(55-57) or bright visible luminescence with quantum yields approaching 100%.(37, 40, 58, 59) Furthermore, pre-synthesized inorganic NCs can be transformed in many ways by post-synthesis exchange reactions (i.e., ligand-exchange(60-63) as well as cation-(64-69) or anion-exchange(70-74)), which provide therefore means for tuning their chemical and physical properties. When these NCs are deposited as a thin film on a substrate, they can serve as active layer for different devices such as solar cells,(75-80) light-emitting diodes (LEDs)(81-85) or biosensors.(86, 87) A more widespread use of colloidal NCs can be achieved when these active materials are deposited not as homogeneous continuous films but only in specific locations of the substrate. Indeed, it was shown in the late 1990s that such patterning could be used for the fabrication of electrical circuits from metallic NCs.(88-96) Later, in the early 2010s, patterning fluorescent QD films was used for detection of biological analytes.(97, 98)These recent examples show the technological interest in using patterned NC films combining the unique properties of bottom-up colloidally synthesized materials with the large-scale versatility of top-down fabrication tools such as lithography (e-beam and others). In order to prepare patterned NC films from a colloidal dispersion one can specifically deposit them only in the desired areas of the substrate (e.g., by microdroplet inkjet printing or microcontact printing).(99-104) However, the spatial resolution that can be achieved by selective deposition is limited to the microscale. Another approach to achieve a more precise localization of colloidal NCs on a substrate at the nanoscale is based on the self-assembly of these NCs on predefined spots or areas which can act as specific anchoring sites for the NCs through different interactions such as capillary forces, (bio)chemical surface binding or electromagnetic forces (magnetophoresis, dielectrophoresis, or plasmonic tweezers).(105-112) All of these approaches, however, require the substrate to be pre-patterned topographically and/or chemically and therefore confine the NC film patterning to these pre-defined areas. Eventually, direct-write methods have been developed for patterning NC films. This last case is the topic of this Perspective. In this approach, the substrate does not need to be pre-patterned (see Figure 1a). Indeed, starting from a continuous homogeneous film of NCs (Figure 1a-i), it is possible to define a pattern directly on the NC film through irradiations with different types of sources (e-beam, X-rays, or UV; Figure 1a-ii). In this case, the film itself works as a sort of negative resist, since not irradiated NCs can be consequently washed away in solvents (“lift-off”) while irradiated NCs remain “glued” to the substrate (Figure 1a-iii). Very recently,(113-115) we have shown that this process can also be used to tune the chemical composition and subsequent properties of irradiated and non-irradiated NCs, which allows for more complex patterning of NC films (Figure 1a-iv). This latest finding, which has already been applied to different QDs (chalcogenide and halide perovskite NCs) represents a major breakthrough in this approach, as it allows the modification of the NCs themselves, opening thus a new set of possible applications in thin film nanofabrication that was until now impossible to obtain through other processes.(116) In this Perspective we will discuss the process of direct lithography on NC films. First we will clarify the physicochemical transformations that occur upon irradiation and enable NCs to remain attached to the substrate and/or inhibit further transformations. We will then review the results that have been reported to date using this approach and analyze the performances in terms of spatial resolution and physical properties of patterned films. Eventually, we will discuss potential future outcomes of this still incipient technique, pinpointing the major actual bottlenecks for a widespread use of this approach in nanofabrication and suggesting different means to overcome them.

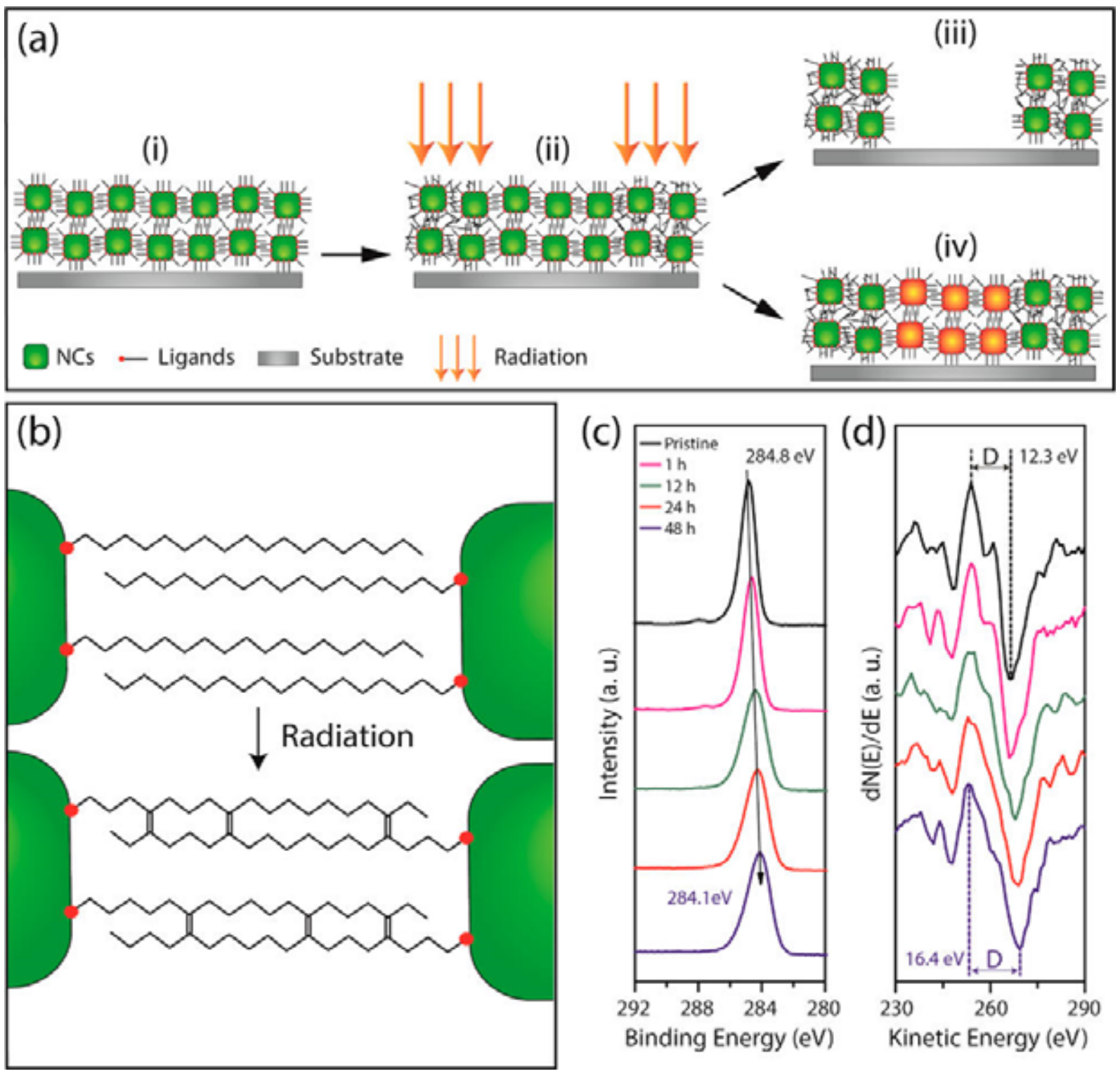

Figure 1: (a) Schematic representation of irradiation-induced patterning of colloidal NC films. (i) Inorganic NCs with organic ligands at the surface are synthesized by a colloidal approach in solution and deposited on a substrate (e.g., by spin-coating). (ii) Selected regions of the NC film are irradiated in order to (iii) selectively fix the exposed regions to the substrate while non-exposed regions are redispersed (i.e., lifted-off) or (iv) selectively modify the composition (and hence physicochemical properties) of non-exposed NCs. (b) Schematic representation of irradiation-induced dehydrogenation and consecutive C=C cross-linking. (c) Evolution of the C 1s XPS spectrum and (d) D-parameter computed from the differentiated C KLL Auger spectrum of organic ligands upon C=C bond formation. Reproduced with permission.(115)

## Fundamentals of Direct-Write Lithography on NC Films

### Intermolecular C═C Cross-Linking of Surface Ligands

Inorganic NCs can be stabilized colloidally either by means of surface charge as described by the DLVO theory (117, 118) or, more commonly in organic solvents by steric repulsion. For this purpose, NCs are usually capped with long linear organic ligands. Indeed, when two NCs passivated with organic ligands come in contact in a solvent, steric hindrance between the two passivating layers results in a repulsive force between the NCs. In the absence of these molecular ligands, the NCs may aggregate to minimize their free surface and eventually cluster into larger objects that become insoluble. When NCs are deposited on a substrate, they obviously retain their organic ligands at the surface, so that, in most cases, the individual NCs that are forming the film can be redispersed in solution if the film is immersed in a good solvent. Reetz et al.(96)discovered that after e-beam irradiation on an NC film, the particles remained “glued” to the substrate. Therefore, they concluded that e-beam irradiation had removed surface ligands and caused the consecutive aggregation of NCs in the exposed areas. The same mechanism was claimed in later reports of e-beam irradiated NC films by other groups.(88-90, 92, 95, 96) However, Werts et al.(93) questioned that ligand stripping was the driving force for NCs insolubility after irradiation. There is now extensive experimental and theoretical evidence that ligand *cross-linking* between adjacent NCs instead of ligand stripping is the main factor driving to the aggregation of the NCs and their apparent anchoring to the substrate. First of all, Werts et al. noted that the irradiation-induced anchoring was more efficient on NCs capped with longer molecules (dodecanethiol, DDT, C12) than shorter ones (hexanethiol, HT, C6). This is consistent with the fact that anchoring occurs through cross-linking of the ligands (which should have higher probability of occurrence on longer molecules), and not through their removal (which would happen preferentially on shorter molecules). Additionally, characterization of the film after exposure and immersion in good solvent showed infrared signal of the ligands and no reduction in film thickness, which suggests that the ligands were not stripped-off. Instead, Werts et al. suggested that, in analogy to previous observations on e-beam irradiation of hydrocarbon polymers and self-assembled monolayers of organosilanes on silica,(119, 120) irradiation induces a cleavage of C—H bonds (dehydrogenation) and the formation of new C═C bonds, as schematically shown in Figure 1b. As these bonds covalently link molecules that are originally separate from each other, the process can be rightly referred to as "cross-linking", although it is worth noting that usual polymer cross-linking involves, on the contrary, the breaking of C═C ($sp^2$) double bonds and formation of new C—C ($sp^3$) bonds. As a consequence, adjacent NCs in the film are chemically bonded and become insoluble in solvents.

The explanation presented by Werts et al. has been corroborated by later works (94, 97, 98, 113, 115, 116, 121-123) to the extent that they rule out the initial hypothesis of ligand stripping. Recently,(113, 115) we have shown by *ex situ* and *in situ* X-ray photoelectron spectroscopy (XPS) and X-ray-excited Auger electron spectroscopy (XAES) that carbon hybridization of capping ligands on an NC film goes from $sp^3$ to $sp^2$ (see Figure 1b–d), meaning that C–H bonds are cleaved and C═C bonds are formed. This partial change in the hybridization of carbon molecular orbitals can be evaluated by the shift to lower binding energies of the C 1s XPS peak and by the evolution of the so-called *D*-parameter (see Figure 1c,d). The *D*-parameter represents the difference in eV between the maximum and minimum of the differentiated C KLL carbon Auger spectrum and evolves linearly with the $sp^2/sp^3$ ratio, from ca. 12 eV for diamond to 21 eV for graphite.(124)

**Primary Beam or Secondary Processes**

Previous work on irradiation of self-assembled monolayers (SAMs) of alkanesilanes on planar oxide surfaces concluded that secondary electrons generated in the substrate cause the cleavage of C–H bonds and subsequent formation of C═C bonds, as described above.(119) A major argument supporting the role of secondary electrons is that the process also takes place when the SAM is exposed to X-rays, in the absence of a primary electron beam. As we have shown, cross-linking of ligands on NC films also occurs upon exposure to X-rays.(114-116) Therefore, also in the case of NC films we can suggest that secondary electrons play a role in the formation of intermolecular C═C bonds. Nonetheless, the evidence that this process takes place also under X-ray irradiation in the absence of a primary electron beam does not rule out that under e-beam lithography the primary beam could also induce the formation of C═C bonds. Although it is not easy to disentangle the role of the primary and secondary electrons in the process, Bedson et al.(90, 91) demonstrated, by varying the type of substrate, that secondary electrons play a major role in the cross-linking mechanism (see Figure 2).

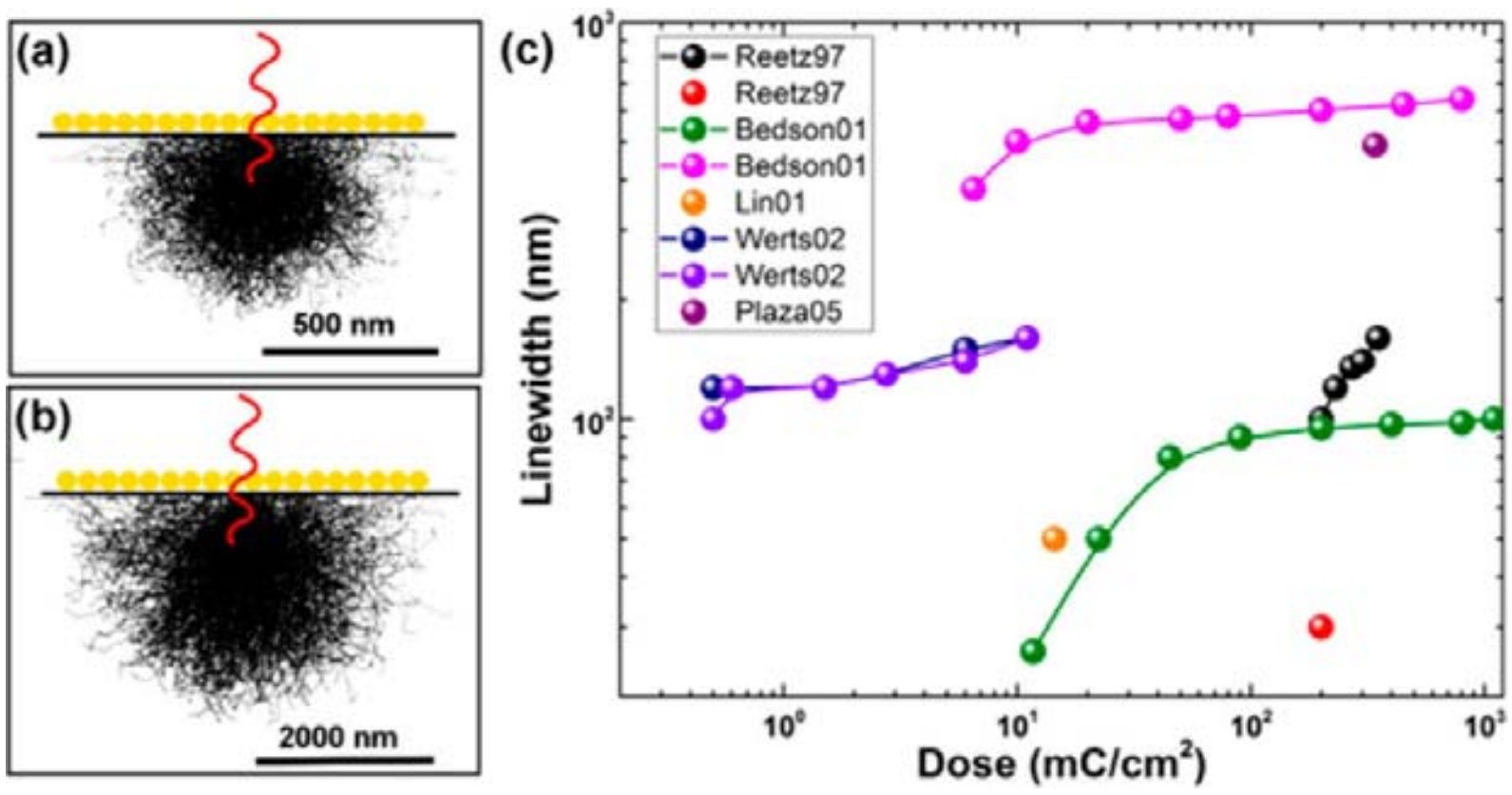

Figure 2: Effect of the substrate showing that secondary electrons are primarily responsible for the lithography process on monolayer NC films. (a,b) Numerical simulations showing secondary electrons generated in two different substrates. Reproduced with permission from ref 91. Copyright 2001 AIP Publishing LLC. (c) Line width vs electron dose obtained for different metal colloids films. The legend refers to first author and year of publication. When the same legend appears for different markers, it refers to different systems (ligands, film thickness, and/or substrate).

In order to demonstrate this point, they deposited thin monolayers of gold NCs on top of silicon wafers with either a 65-nm thin thermal $SiO_2$ layer or a several-micrometers-thick sputtered $SiO_2$ layer. As revealed by numerical simulations (see Figure 2a,b), the secondary electron emission yield is much higher in the latter case (Figure 2b) than in the former, meaning that regions farther away from the primary beam “spot” are irradiated by secondary electrons (Figure 2a; note that the scale is different in both representations). This fact explains that at equal (primary) electron dose, the line width obtained on sputtered $SiO_2$ is several times larger than on the thin thermal $SiO_2$ substrate (see Figure 2c). Furthermore, when they conducted the same experiments with a 135-nm-thick multilayer of gold NCs, the influence of the underlying substrate was minimized, as most of the secondary electrons came from the NC layer itself in this case. Thus, it is evident that secondary electrons are the main source of ligand C═C cross-linking and that, for thin NC films, the underlying substrate may play a crucial role in the achievable resolution.

**Evolution of Materials’ Properties upon Irradiation**

We have previously shown that irradiation induces chemical changes in the ligands that cap the inorganic NCs. The question remains though as to whether the NC themselves are affected by the irradiation. Furthermore, it is interesting to evaluate the properties of the final material that results from the irradiation, that is the network of inorganic NCs partially bonded through their ligands. Another way of seeing it is as a carbonaceous matrix embedding inorganic NCs.

Initially, direct lithography on metallic NC films was thought to lead to the agglomeration of the NCs due to partial fusing/sintering of the cores.(96) However, later work excluded this hypothesis, as already discussed.(92) In fact, neither the size nor the crystallinity or chemical composition of inorganic NC cores seemed to be affected by the lithography process.(115) The fact that the lithography process only affects the ligands directly does not mean, though, that the overall film properties are identical before and after irradiation. The first noticeable modification of the film properties which has been the obvious reason to develop this methodology is the insolubility of the exposed regions. Although irradiated NCs appear to be “glued” to the substrate, it would be more accurate to say that irradiated NCs are “glued” to each other, forming large objects that are thus insoluble in solvents that would otherwise redisperse individual (non-linked) NCs.

For irradiated metallic NCs, an important property to evaluate on the final material is its electrical conductivity. All the different published works dealing with the formation of metallic nanowires from colloidal metallic NCs agree that the final conductivity of the so-formed wires is orders of magnitude lower than the bulk metal counterparts (details are given in the next section), which is consistent with the fact that metallic NCs are not sintered but rather embedded in a mainly non-conductive carbon matrix. It could be thought that the partial C═C cross-linking of this carbon matrix should at least improve the conductivity with regards to the non-irradiated NC film. However, there is to the best of our knowledge no proof that the irradiation-induced cross-linking of the ligands substantially improves the film conductivity. When dealing with fluorescent QD films, another important aspect to consider is the evolution of the photoluminescence quantum yield (PLQY) upon irradiation. As will be further detailed in the next section, the PL intensity of the film drops with exposure dose and, as has been shown for films of strongly fluorescent cadmium chalcogenide or halide perovskite NCs, the drop in PL intensity could be drastic and reach almost a total quenching.(97, 115) The exact mechanism leading to fluorescence quenching has not been fully elucidated to date. Nonetheless, it is known that the PLQY of QDs is affected by the quality of the surface.(125) Therefore, it is reasonable to suggest that irradiation, which as previously discussed causes chemical modifications on the NCs surface, may lead to the formation of surface trap states. These traps then can act as non-radiative recombination centers for excitons created at the QD cores, thus quenching the photoluminescence of the film. Eventually, we have recently shown that the partial ligand C═C cross-linking acts as a very efficient barrier against several external agents such as cations,(113) anions,(115, 116) water(115) or short chain alkylamines.(114) These atomic or molecular species which could penetrate pristine films and alter the NC cores are instead blocked by the irradiation-induced ligand cross-linking. The exact mechanism leading to this unprecedented shielding has not been elucidated yet.

Overall, the data currently available suggest that, upon irradiation by e-beam or X-rays, inorganic NC (cores) are not significantly affected, retaining their size, morphology, and crystallinity. However, the cross-linking of surface ligands (shells) does affect the global properties of the film (e.g., insolubility, photoluminescence quenching, shielding against external agents). Further fundamental studies are needed for a better understanding of such “hybrid” films, which are essentially different both from films of individual NCs (non-irradiated) or continuous bulk films (no ligands). On this regard, it would be especially interesting to provide a theoretical background to the recently evidenced impermeability of the irradiated films toward numerous small molecular species.

## Main Achievements of Direct-Write Lithography on NC Films

### Metallic Nanowires

Gold,(88-94) palladium,(95, 96) and bimetallic palladium–platinum(96) NCs of sizes around 2–5 nm have been used to create metallic wires by direct e-beam lithography. In this application, the main characteristics sought after are the smallest possible spatial resolution (line width) and the best metallic behavior (conductivity). Figure 2c shows the line widths that have been reported at different (primary) electron doses. For a given film (identified by identical markers), the line width decreases with electron dose up to a threshold value below which the film is no longer “glued” to the substrate. This behavior is similar to that of conventional lithography on polymer resists. The threshold dose defines the sensitivity of the resist, or, in this case, the NC film. Reetz et al.(96)found a threshold dose around 200 $mC/cm^2$, whereas further work by Bedson et al.(90, 91)achieved a value 1 order of magnitude lower at 11.7 $mC/cm^2$. Eventually, Werts et al.(93) found a sensitivity as low as 0.5 $mC/cm^2$ on films of gold NCs, which is of the order of magnitude of conventional e-beam lithography resists such as PMMA (0.05–0.5 $mC/cm^2$ depending on equipment, according to MicroChem datasheet).(126) The discrepancies found between these values are related to the length of the molecular ligands, the film thickness, and the nature of the underlying substrate. Werts et al.(93) showed that longer molecular ligands at the surface of the NCs lead to lower threshold doses. The reason for this observation has to do with the crucial role of the organic ligands in the lithography process, as explained in the previous section. The impact of the film thickness and the underlying substrate can be seen when comparing both series reported by Bedson et al.(91) (green and pink markers) and results reported by Reetz et al.(96)(black markers). Bedson et al. used monolayer films and the line width vs dose response saturated in both cases above 10–50 $mC/cm^2$, whereas Reetz et al. used thick multilayer films of 180 nm and got a linear dependence of the line width vs dose in the 200–300 $mC/cm^2$ range (and no writable feature below 200 $mC/cm^2$). On the other hand, the only difference in both cases presented by Bedson et al. resides in the underlying substrate and has a dramatic effect on the line width (about 1 order of magnitude higher for monolayer films deposited on sputtered $SiO_2$, pink triangles, than for the same monolayer deposited on thermal $SiO_2$, blue markers). These dependencies on film thickness and underlying substrate are related to the effect of secondary electrons, as previously explained. As a consequence the reported line widths in different works varies greatly as it depends on multiple factors (NC material, film thickness, substrate, electron dose, and nature and number of ligands, indirectly related to NC dimensions). In the last part of this Perspective, we will discuss on the ultimate achievable resolution and its limiting factors. It is noteworthy nonetheless that the early work by Reetz et al. in 1997 already demonstrated 30 nm line widths, a value that has only slightly been improved by Bedson et al. in 2001 (26 nm). These values are just above the current achievable resolution by conventional e-beam lithography and anyhow suitable for many electronic applications. Therefore, one could expect a widespread use of this methodology for the fabrication of nanoelectronic devices, assuming that the electrical performances (conductivity) are adequate. The first reports from 1997–1998(88, 89, 95, 96)showed that the so-formed wires on gold or palladium NCs exhibit a metallic behavior with linear *I*–*V* curves and temperature-dependent resistivities. Nonetheless the resistivity values obtained in this approach were around 2 orders of magnitude higher than for bulk metals.(96) This poor conductivity was related to the presence of carbon in the film, and it was found that performing a subsequent annealing step could improve the conductivity, which still remained well below the values of carbon-free bulk metal.(95, 96) These observations of poor conductivity due to the presence of carbon in the nanowires still hold in following reports by Plaza et al.(94) On the other hand, Bhuvana et al.(123) showed in 2008 that it was possible, using a continuous organometallic resist made of palladium hexadecylthiolate, to write 30-nm-wide lines at a dose as low as 0.135 $mC/cm^2$ and yielding a conductivity close to that of bulk palladium.

It appears therefore that the application of direct-write lithography on colloidal metallic NC films is hindered mainly by the poor conductivity that can be achieved (below the values of bulk metals but also of wires defined on continuous organometallic films) and to a minor extent by the relatively higher electron doses that are needed. However, the work developed on metallic colloids has been of great value to understand the different radiation–matter interactions that occur in the lithography process on NCs (see first main section of this Perspective, Fundamentals of Direct-Write Lithography on NC Films). Eventually, this knowledge has been applied for the direct-writing on films of semiconductor quantum dots as detailed hereafter.

**Patterned QD Films**

QDs have several advantages in comparison to their bulk semiconductor counterparts due to the intrinsic properties linked to the nanoscale (i.e., quantum confinement) as will be detailed hereafter. Therefore, the ability to pattern semiconductor NC films represents a real advantage with respect to patterning continuous semiconductor films. In 2011 the group of Rotello applied the methodology developed on metallic NC films to films of fluorescent semiconductor core–shell CdSe/ZnS quantum dots capped with trioctyl phosphine oxide (TOPO).(97) A 55-nm-thick film of QDs was spin-coated on a gold substrate and microsquares were exposed to electron beam at doses ranging from 100 to 8000 $\mu C/cm^2$. As can be seen in Figure 3a,b, the fluorescence decreased after exposure to the beam, although it was not fully quenched even at the highest dose. The authors found no significant change in the photoluminescence decay times before and after exposure. After washing with toluene, all the exposed regions (even at the lowest dose of 0.1 $mC/cm^2$) remained bound to the substrate.

Fluorescent QD patterns defined by direct e-beam lithography have been used, with appropriate surface functionalization, for the recognition and detection of different biological analytes such as proteins(97, 98, 121)or cells.(98) It is interesting to note that Palankar et al.(98) could write micrometric features on a film of PEG-functionalized QDs with a dose as low as 0.01 $mC/cm^2$. These results from 2013 by two different groups demonstrate the potential of the direct-write lithography process on QD films for applications in biosensing.

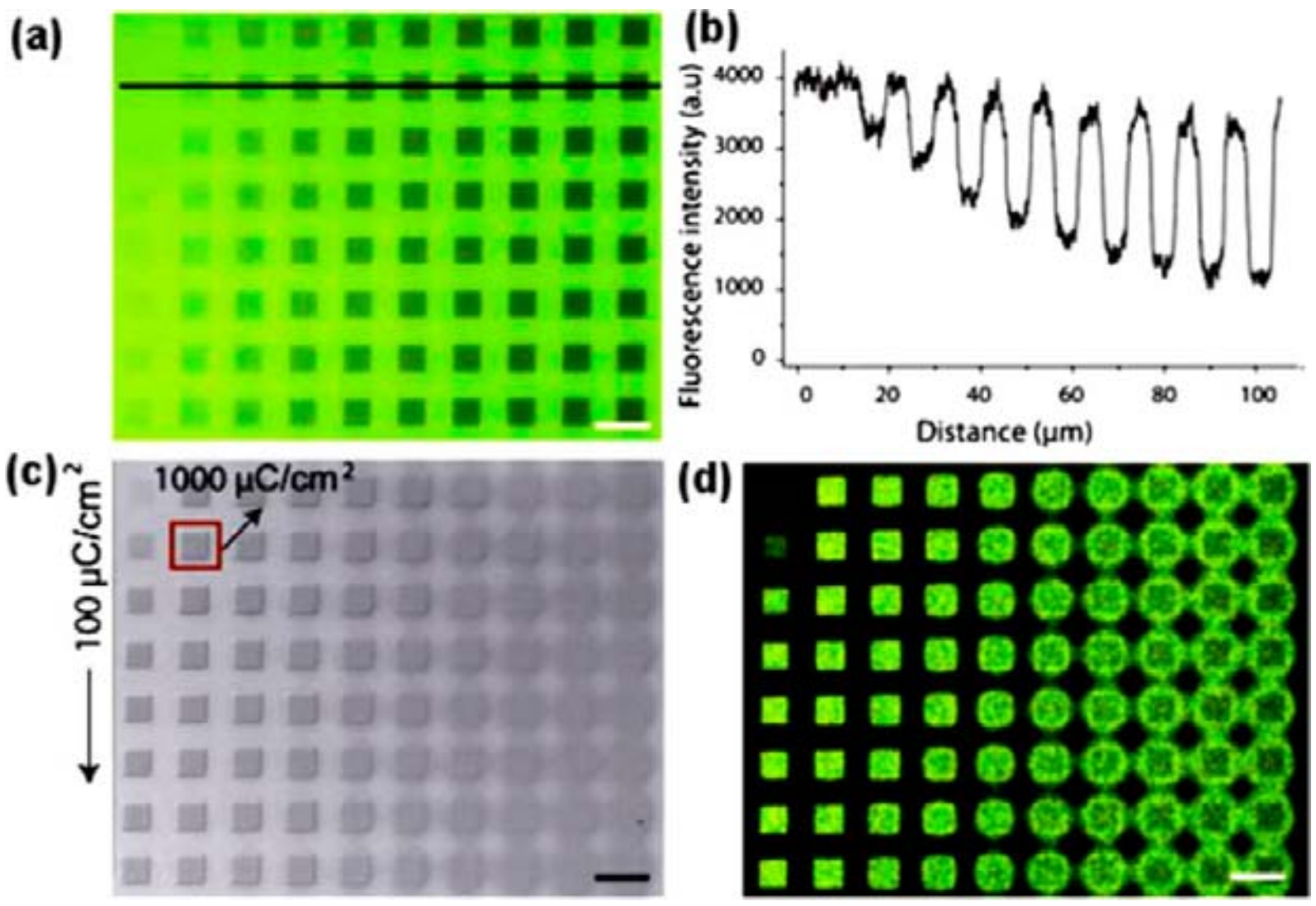

Figure 3: (a) Fluorescent image of QD test pattern before washing with toluene. (b) Fluorescence intensity across the patterns in panel (a). (c) Bright field and (d) fluorescent image of QD test pattern after washing with toluene. The dose was varied between 100 and 8000 μC/cm$^2$. The scale bar is 10 mm. Reproduced with permission from ref 97. Copyright 2011 The Royal Society of Chemistry

**Beyond Simple Patterning: Tuning NCs' Chemical Reactivity**

Irradiating selected regions of a NC film can be used not only to create patterns on the film itself by redispersion of non-exposed areas, but also to induce further modifications of the underlying substrate or the film itself, yielding more complex patterns. As an example, Hogg et al.(122) used a patterned film of iron oxide NCs defined by direct e-beam lithography as a hard mask for a subsequent etching step of the substrate. Their approach involves e-beam irradiation ("curing"), followed by $O_2$ plasma to partially remove ligands and $CF_4$-mediated etching (see Figure 4). They found out that the "curing" step was crucial to avoid particle aggregation and film-cracking, thus leading to a finer resolution and greater pattern fidelity in the etching process.

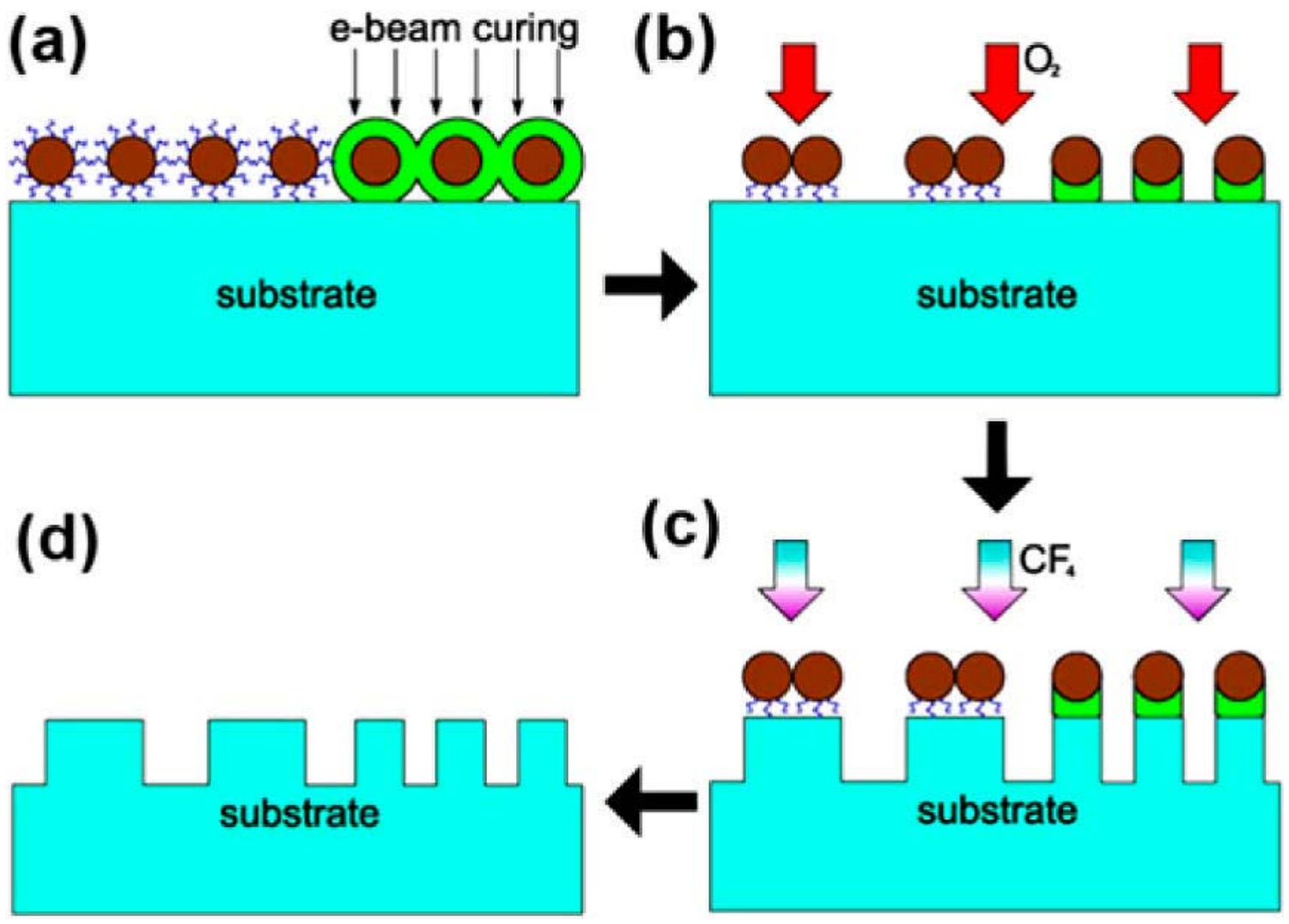

Figure 4: Irradiated NC films used as hard mask for substrate etching. Non-irradiated NCs aggregate hindering spatial resolution. Reproduced from ref.122 with permission from IEEE.

Recently, we have shown in our group that direct lithography on QD films either with electron beams or X-rays leads not only to an enhanced adhesion to the substrate but also makes the exposed regions less prone to undergo various chemical transformations, as they become partially sealed off from the external environment.(113-115) For instance, we demonstrated that pristine TOPO-capped CdSe/CdS NC films can be transformed to $Cu_2Se/Cu_2S$ by cation-exchange reactions with copper precursors. Note that, in this case, these precursors were dissolved in solvents in which the NCs themselves were not dispersible, otherwise the film would have been damaged. However, when some regions of the film are previously irradiated by e-beam or X-rays, they become refractory to cation exchange, so that when the whole film, after irradiation, is exposed to a solution containing $Cu^+$ species, it develops into a patterned film of cadmium and copper chalcogenide NCs (see Figure 5a–d). This strategy was used by us to define luminescent patterns of cadmium chalcogenides on a film of non-luminescent copper chalcogenides, or even conducting wires of copper chalcogenides in a non-conducting film of cadmium chalcogenides.(113)

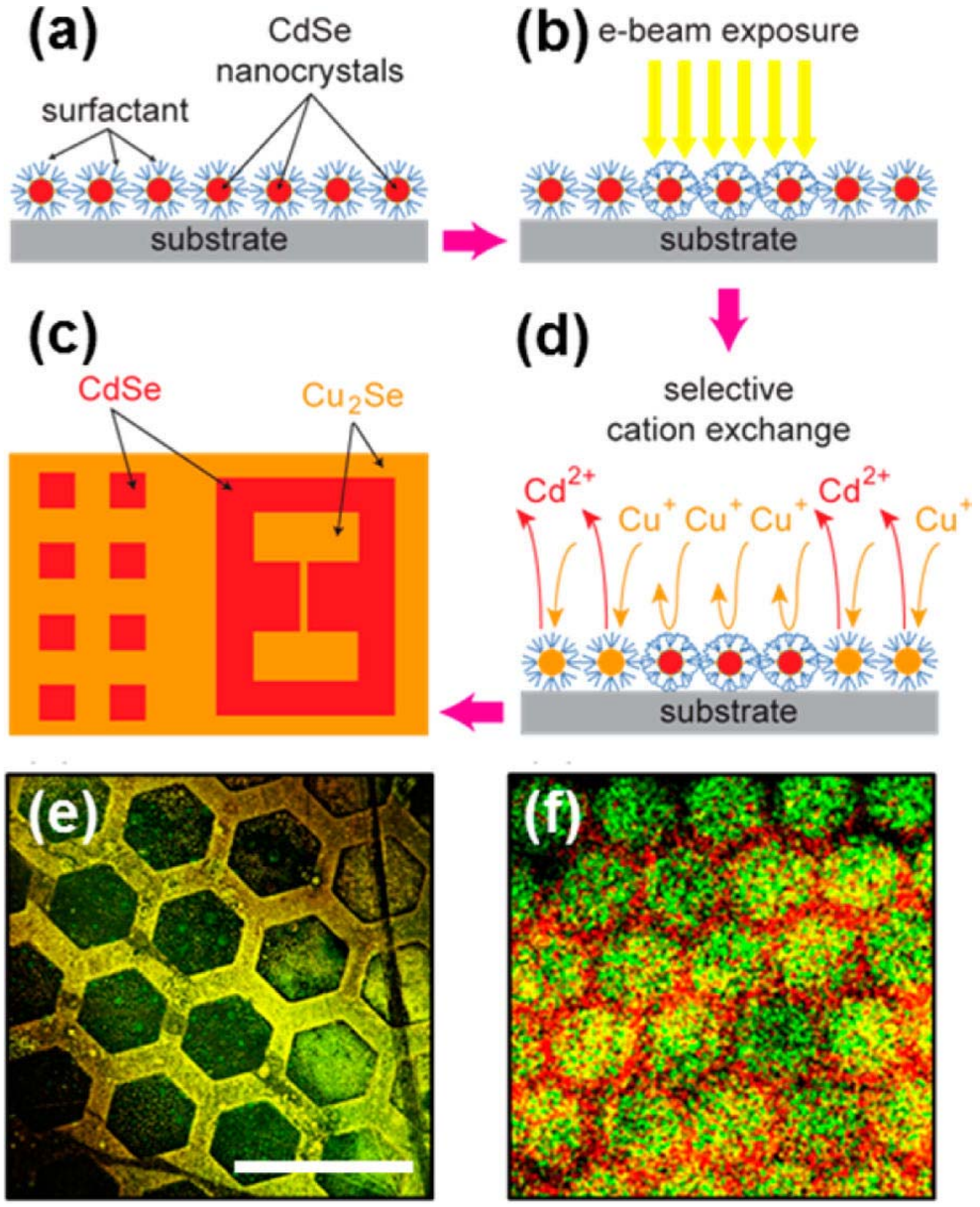

Figure 5: (a−d) Selective cation exchange. Reproduced with permission from ref 113. Copyright 2014 American Chemical Society. (e) PL and (f) chemical maps of $CsPbBr_3/CsPbI_3$ films by X-ray lithography. Reproduced with permission from ref 115. Copyright 2017 American Chemical Society

We also developed a similar approach on inorganic halide perovskite NCs. Cesium lead halide ($CsPbX_3$; X = I, Br or Cl) perovskite NCs have gathered high interest in the past 2–3 years owing to their high photoluminescence quantum yield and emission tunability throughout the visible spectrum, which can be easily achieved by post-synthesis anion-exchange reactions.(37, 70, 71)Similar to the cation-exchange inhibition on cadmium chalcogenides, we have demonstrated that the exposure to e-beams and X-rays under vacuum inhibits anion exchange reactions on cesium lead halide NC films.(115) This allows the patterning of luminescent films at different wavelengths (see Figure 5e,f) and even the fabrication of white-light-emitting films.(116) In another approach,(114) we used irradiation-induced ligand cross-linking to inhibit further reactions with volatile amines that, for non-protected $CsPbBr_3$ NCs, leads instead to structural transformation into $PbBr_2$-depleted $Cs_4PbBr_6$ NCs.(114, 127) Eventually, we also noted that irradiated films of $CsPbI_3$ NCs showed an enhanced resistance toward reactions with oxygen and water, which otherwise degrade the NCs.(115) These results show that direct lithography on NC films is a technique that goes well beyond "simple patterning" (i.e., maintaining the original NCs in the exposed areas while redispersing the rest in a solvent) and that it can in turn be used to fine-tune the chemical composition and hence the optoelectronic properties of different regions of the film.

## Perspectives of Direct-Write Lithography on NC Films

### Ultimate Resolution: Few NCs, Single NC, or Fraction of NC?

As direct-write lithography on NC films operates through the cross-linking of surface ligands between adjacent NCs, it could be thought that the final achievable resolution cannot be as small as a single NC. Indeed, although inter or intramolecular C=C bonding between ligands of a single NC can occur, there is no reason why such NC would become insoluble in a solvent that would redisperse non-irradiated NCs. Nonetheless, a workaround for single NC patterning can be proposed through substrate functionalization (see Figure 6a).

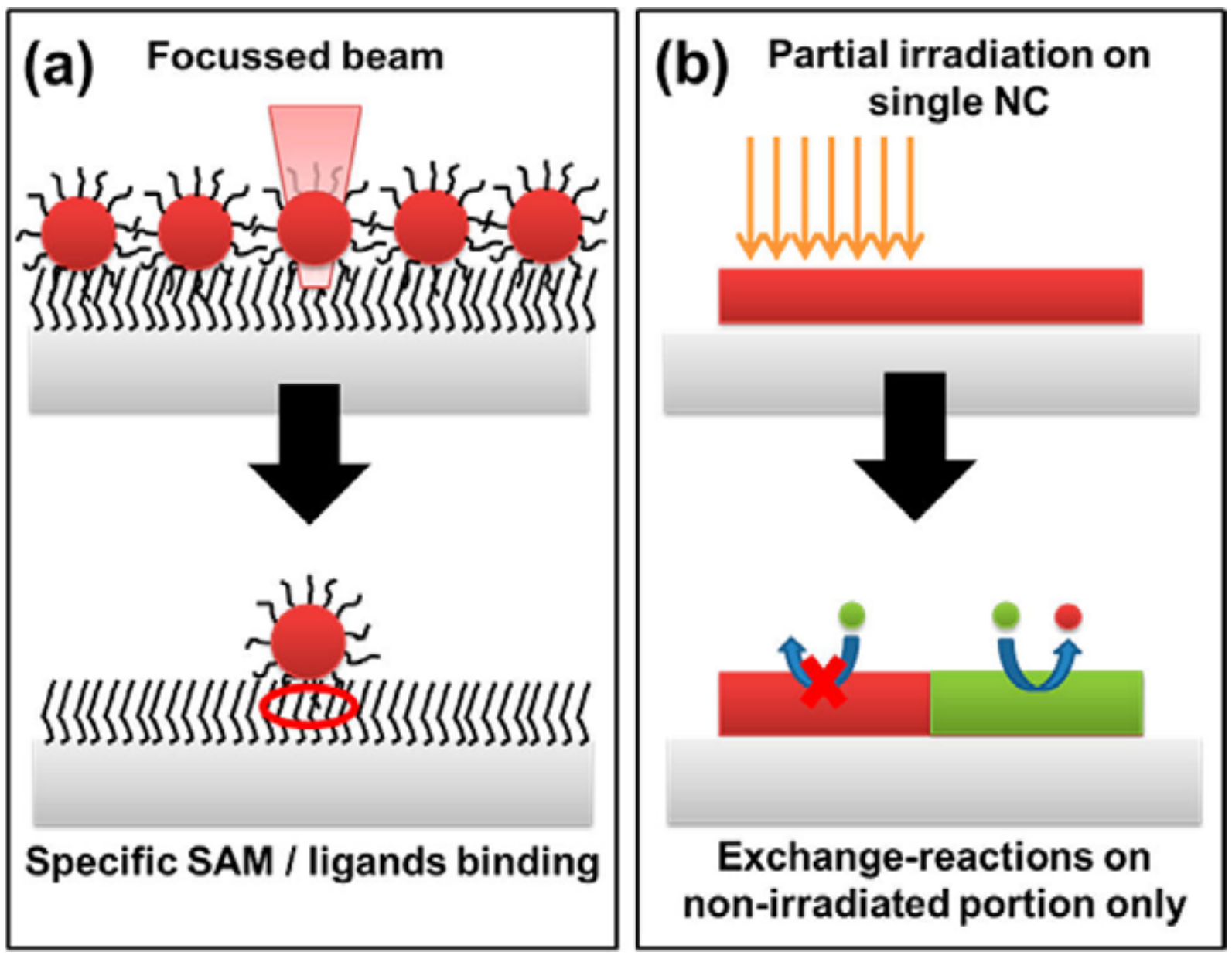

Figure 6: (a) Schematic representation of possible single-NC patterning through specific cross-linking of surface ligands with functionalized substrate. (b) Schematic representation of possible heterostructure formation through ligand C=C cross-linking of a fraction of NC.

Self-assembled monolayers (SAMs) of different organic molecules have been attached on different planar substrates and even used for the specific binding of colloids. In this case, the interaction between the substrate (SAM) and the NC (ligands) should be weak and non-specific and would only become specific under irradiation. By simple analogy with the process described so far for inter-particle bonding, a simple alkane molecule for the SAM should be useful for C═C bonding with the NC upon irradiation, while the non-irradiated regions could be “lifted-off”. In this case, the direct lithography process would really fix the NCs to the substrate. This methodology is somehow already present in the work of Palankar et al.(98) although the role of particle/substrate vs particle/particle binding is not clearly defined and the aim of that work was not to attain the finest possible spatial resolution. To achieve single NC arrays with this method, one should furthermore ensure that, upon irradiation, only NC/substrate bonds are created and not NC/NC bonds. This might not be simple to achieve with ligands and SAMs of the same nature (alkane chains). An option is to focus the writing beam below the size of a single NC as presented in Figure 6a. Writing with near field probes like an electrically biased AFM tip could represent an interesting strategy in this process.

If the final goal of the lithography process is not to make irradiated regions insoluble but to enable or inhibit further chemical transformations as partial replacement of cations or anions, then the final resolution that could be achieved could even reach the fraction of a NC. As ion diffusion might be dependent on crystallographic directions, it should be possible to pattern a single nanowire. In principle, if only a fraction of the nanowire is irradiated, cation or anion-exchange reactions should only affect the non-irradiated fraction, leading thus to the formation of a heterostructure (see Figure 6b). Indeed, partial exchange reactions leading to segmented nanowires have already been achieved with traditional masking (using a polymeric resist).(128, 129) This could be the basis for novel devices as presented hereafter.

**Toward Ultraviolet Lithography**

Whether it is for simple patterning by redispersion of non-irradiated regions or for inhibiting further transformation of the NC cores, all the reported work so far is based on the cleavage of C–H bonds of alkane ligands and consecutive intermolecular formation of C═C bonds. This has been achieved mainly through e-beam lithography with typical acceleration voltages of few tens of kV in vacuum. As we have recently shown, this can also be achieved upon irradiation of X-ray photons, also in ultrahigh vacuum. In contrast, to achieve similar results with UV light remains an important challenge. Indeed, Clarke et al. noted that “attempts were made to pattern the material using 254 nm UV lithography, but it was found to be insensitive to this wavelength.”(89)

Having “writable” NC films under standard photolithography setups (in air, with excitation sources around a few hundred nanometers in wavelength, near UV) would greatly enhance the appeal of direct lithography on NC films. Indeed, e-beam and X-ray lithography are much more time-consuming and demanding in terms of constraints (e.g., ultrahigh vacuum) than standard photolithography. In order to achieve photolithography on NC films one may think of replacing the alkane ligands by UV-sensitive ones, such as photopolymerizable organic or inorganic molecules, either directly during the synthesis of by post-synthesis ligand-exchange.

Ligand-exchange on NCs can be performed in solution (before deposition on a substrate) or in films (after deposition) and is quite a standard practice.(60-63) As an example, Alloisio et al.(130) performed a direct synthesis of gold NCs coated with diacetylene henicosa-10,12-diyn-1-yl (DS9) disulfide. This ligand can polymerize under UV radiation at 254 nm. In their work Alloisio et al. demonstrate intra-particle polymerization in dilute toluene dispersions of NCs. However, it is reasonable to assume that if a similar irradiation were carried out on dense films of NCs deposited on a substrate, the diacetylene ligands would not only cross-link between molecules of the same NC (intra-particle) but also with ligands of adjacent NCs in contact (inter-particle), leading thus to similar results as those described in this Perspective. In a recent report that was published during the peer review process of this Perspective, Wang et al. demonstrated for the first time photolithography on inorganic NC films capped with different surface ligands.(131)

**Devices**

The nanofabrication possibilities opened by direct writing on NC films, especially by the selective modification (e.g., through anion or cation-exchange reactions) of the NCs themselves, can be used to design novel optoelectronic devices. For instance, selected regions of a film of conductive NCs (for example, $Cu_{2-x}E$, E = S, Se, Te) can be exposed either to an electron beam or to an X-ray beam (or to a laser beam). The treatment will make the exposed regions inert to cation exchange. The unexposed regions will remain instead reactive, and therefore can be transformed into regions made of semiconductor NCs, for example if $Cu^{+}$ ions are exchanged with ions such as $Zn^{2+}$, $Cd^{2+}$, $Pb^{2+}$ (Figure 7a).

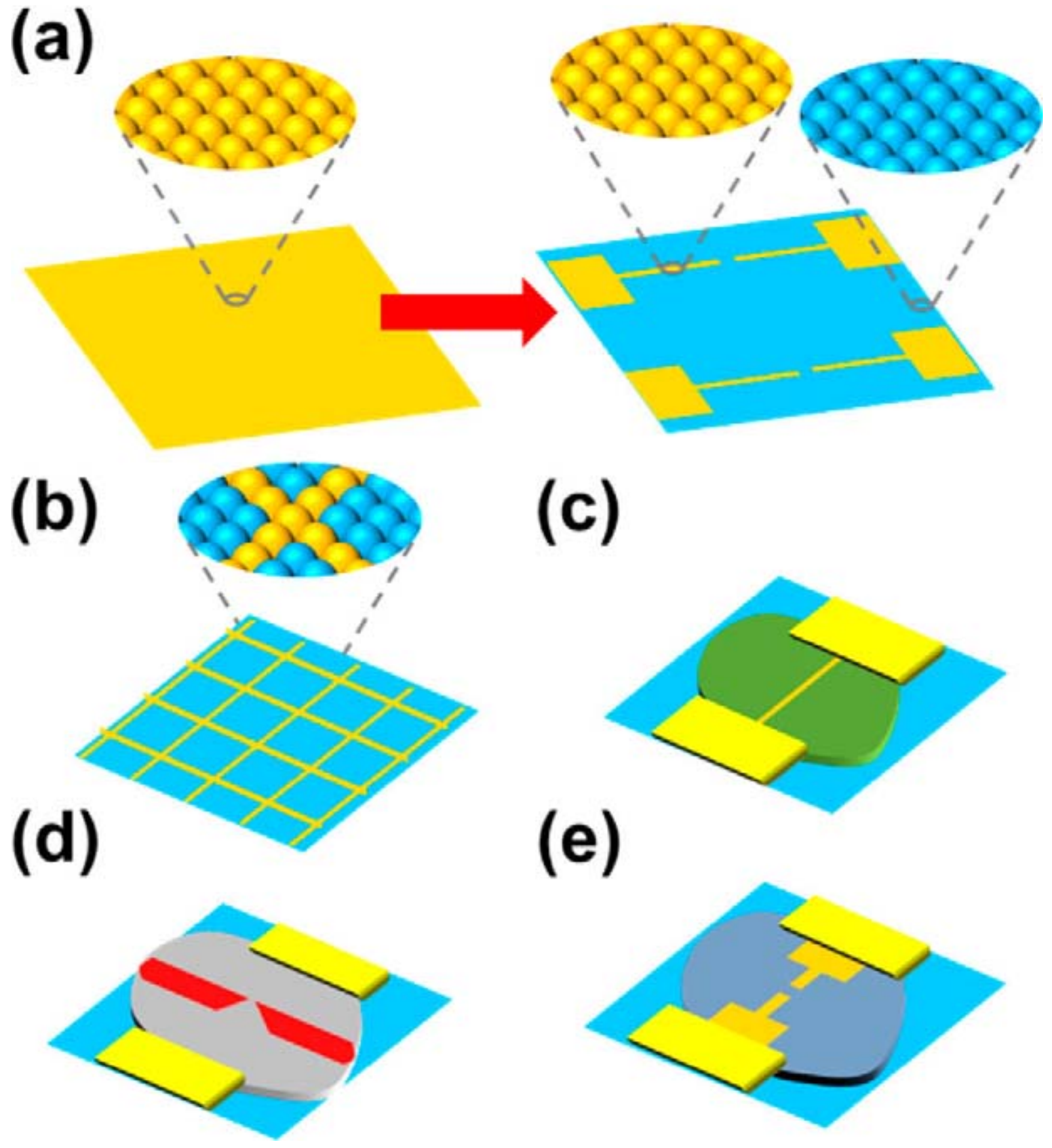

Figure 7: (a) 2D patterns of metallic (yellow)/semiconducting (blue) regions by combining masked cation exchange and atomic ligand passivation. (b) Arrays of individually exchanged NCs within a monolayer film; (c) Electrically addressed

single conductive line within a nanosheet. (d) A quantum point contact and (e) a tunnel barrier drawn in a single nanosheet.

Eventually, stripping the ligands off the surface of NCs in the cation-exchanged regions can improve the film conductivity. In particular, it could be interesting to selectively convert linear arrays of NCs within a 2D monolayer film of NCs (Figure 7b), thus creating conducting single NC chains (NC diameter 5–10 nm), or even a line/pattern of a similar resolution within a nanosheet (Figure 7c). Different patterning geometries and combinations of substrates can be explored in order to prepare basic elements such as a planar photodetector and a field effect transistor (FET).

More advanced structures can be obtained by patterning single nanosheets (Figure 7d,e). For instance, by writing a few-nm-wide line and subsequent cation exchange to a large band gap material, a tunnel junction can be created. If one could reduce the line to a few-nm-sized dot and exchange to a low-gap material, then even a quantum point contact could be realized. The eventual charge transport between the proposed devices and outside contacts is of course equally important. Selective modifications on nanowires can be used to test basic circuit elements such as ohmic and Schottky contacts and pn-junctions. An example of fabrication is the one sketched in Figure 8.

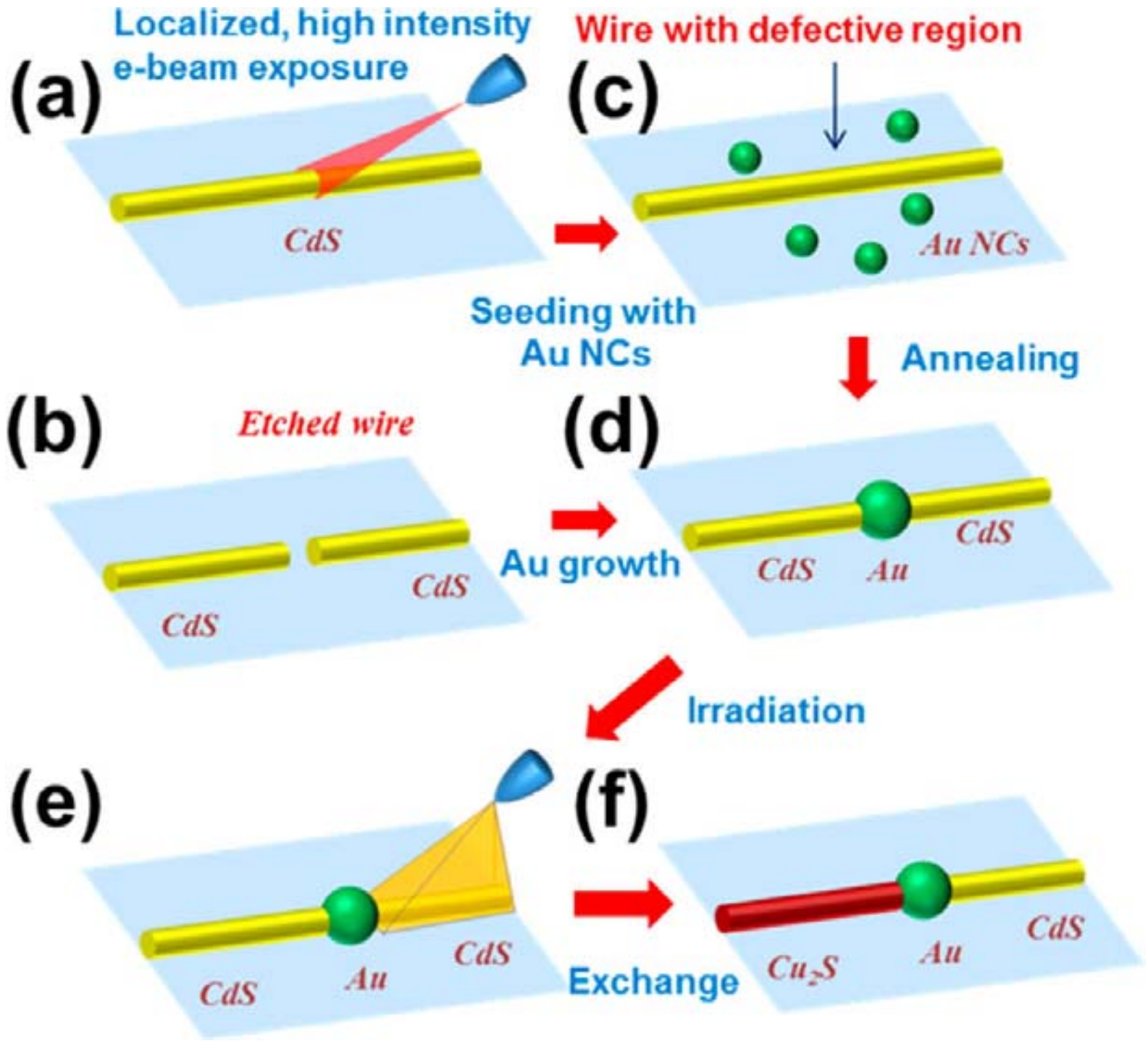

Figure 8: An example of how, in principle, an initial nanowire can be transformed into segments of various materials connected together

Starting from a colloidal semiconductor nanowire on a substrate, a metal segment can be inserted in it or at one tip of the wire. Three possible options exist, in principle, to achieve this: (i) to irradiate with a high-intensity, localized e-beam to create a defect in the wire (Figure 8a); (ii) to have a wire with already a defective region that promotes metal insertion; (iii) to have a wire with one section etched away (see Figure 8b). Starting from these three possible cases, one may seed the treated sample with metal nanoparticles and then subject it to thermal annealing (Figure 8c). Annealing will trigger atomic/cluster diffusion from the metal particles to the reactive regions of the wire (either the defective/irradiated region or the tips facing the etched region) creating a metal domain there (Figure 8d). The process can be further implemented by exposing the semiconductor-metal heterostructure to a more defocused and lower intensity beam, which may locally make the exposed regions refractory to cation exchange (Figure 8e). Then, a last step of cation exchange will affect only the non-exposed regions (Figure 8f). Such asymmetric structures can be expected to show diode-like behavior.

## Conclusions and Outlook

Although direct writing on inorganic nanocrystal films with electron beams was demonstrated two decades ago, the possibilities offered by this technique have not been extensively exploited. The original work focused on defining metallic nanowires on gold or palladium NC films. These reports helped to greatly understand the mechanism of direct lithography on NCs, especially the crucial role of the molecular aliphatic ligands which become cross-linked by covalent C═C bonds resulting from the cleavage of C–H bonds mainly by secondary electrons. The electrical conductivity obtained by this approach was scarce (precisely because of the carbon content coming from the ligands) in comparison to bulk metals which may explain why this method of fabricating (carbon-containing) metallic nanowires did not spread widely. Nonetheless, the knowledge developed on metallic NC films was later implemented to create fluorescent patterns on semiconductor NCs (quantum dots). This renewed the interest in e-beam lithography on NC films, especially for applications in biosensing. Recently, we demonstrated that exposure with e-beams and/or X-rays of QD films can be used not only to define simple patterns (by dissolution of non-exposed regions in solvents) but also to allow or inhibit further chemical transformations on the NC cores of exposed or non-exposed regions. This striking finding, which has now been confirmed several times, still lacks a strong theoretical understanding. In fact, although it is easy to understand that cross-linked NCs become insoluble in good solvents simply because of size considerations (cross-linked NCs can be seen as single “bulky” objects) it is not trivial to understand why this cross-linking would block the diffusion of small species such as single ions, beyond the hand-waving argument of forming a “tighter” barrier at the surface. Further fundamental studies, varying the nature of the ligands (e.g., aromatic ligands instead of aliphatic ones) and the irradiation conditions, coupled with more in-depth characterizations on the stability of these ligands should help to provide a better understanding of the reasons behind this exceptional “passivation” induced by ligand C═C cross-linking. We especially demonstrated that anion- and cation-exchange reactions can be blocked by irradiation of the NCs, enabling thus the patterning of the film with NCs of different optoelectronic properties (e.g., conductive and non-conductive regions or regions fluorescing at different wavelengths). Current developments suggest that such patterning can be realized down to sub-NC resolution, selectively modifying portions of a single NC. These encouraging demonstrations suggest that we have only scratched the surface of what can be achieved by direct lithography on NC films. Indeed, many optoelectronic devices (e.g., FETs or photodetectors) can be designed by the selective transformations of NC films or single NCs, which represents unprecedented miniaturization possibilities and the possibility to fabricate quantum devices with tunnel barriers or quantum point contacts.

In parallel to the quest for miniaturization, if direct lithography on NC films is to become widely used in optoelectronic fabrication, it should be interesting to achieve similar effects as those obtained by e-beams or X-rays with standard photolithography setups. This would allow the patterning of large areas in a less time-consuming way with less constraints (e.g., without need of vacuum). In order to achieve this, we suggest that further developments on the molecular ligands that passivate the NCs should be made. In fact, replacing the standard aliphatic surfactants widely used for inorganic NC synthesis with photopolymerizable organic or inorganic ligands should be useful to achieve UV lithography on NC films.

## Acknowledgment

The research leading to these results has received funding from the seventh European Community Framework Programme under Grant Agreement No. 614897 (ERC Consolidator Grant “TRANS-NANO”) and from Framework Programme for Research and Innovation Horizon 2020 (2014-2020) under the Marie Skłodowska-Curie Grant Agreement COMPASS No. 691185 and under the Marie Skłodowska-Curie grant agreement SONAR No. 734690.